\documentclass[12pt,twoside,bezier]{article}

\def\beq{\begin{equation}}
\def\eeq{\end{equation}}
\def\beqa{\begin{eqnarray}}
\def\eeqa{\end{eqnarray}}

\begin{document}

\thispagestyle{empty}

\title{\large\bf 
A ROUGHENING TRANSITION INDICATED BY 
THE BEHAVIOUR OF GROUND STATES}
\author{R. Koteck\'y and S. Miracle-Sole}
\date{\normalsize 
Dept. of Mathematical Physics, Charles University, 
V. Holesovi{c}k{\'a}ch 2 
18000 - Praha 8 - Czechoslovakia\break 
Centre de Physique Th\'eorique, CNRS
Luminy, Case 907,\break F-13288 Marseille Cedex 9, France}
\maketitle

\centerline{\small\bf ABSTRACT}


\begin{quotation}

Our aim in this contribution is to present some 
illustrations for the claim that already by looking at 
the ground states of classical lattice models, 
one may meet some interesting and non-trivial 
structures.

\end{quotation}


\parindent 2cm
\parskip 14pt

\bigskip

{\sc Keywords:} Roughening transition, Ising spin systems, 
BCC lattice, equilibrium crystal.


First of all we shall clarify what we mean by ground states. 
We consider a finite spin model on a lattice $\cal L$ with 
configuration space $\Omega=S^{\cal L}$ and finite range 
interaction $\{\varphi_\Lambda,\Lambda\subset{\cal L}\}$. 
The energy in a finite volume $V\subset{\cal L}$, under a boundary 
condition (b.c.) ${\bar\sigma}\in\Omega$ is, in an obvious 
notation,
$$
H_V(\sigma_V\mid{\bar\sigma})=\sum_{\sigma_V:V\cap\Lambda\ne\emptyset}
\varphi_\Lambda(\sigma_V\cup{\bar\sigma}_V)
$$ 
The collection of finite volume Gibbs states (a specification) 
$$
\mu^{\beta H}_V(\sigma_V\mid{\bar\sigma})=
Z_V({\bar\sigma})^{-1}\exp\,(-\beta H_V(\sigma_V\mid{\bar\sigma}))
$$
where 
$$
Z_V({\bar\sigma})=
\sum_{\sigma_V}\exp\,(-\beta H_V(\sigma_V\mid{\bar\sigma})
$$
determine (by the DLR equations) the set $G(\beta H)$ of 
the Gibbs measures on $\cal L$ corresponding to a hamiltonian 
$H$ and an inverse temperature $\beta$. 
If a Gibbs state $\mu\in G(\beta H)$ happens to equal the limit 
$$
\mu=\lim_{V\uparrow{\cal L}}\ \mu^{\beta H}_V(\cdot\mid{\bar\sigma})
$$
under a fixed b.c. ${\bar\sigma}$, we shall call it 
$\underline{\hbox{Gibbs state corresponding to a b.c. }
\bar\sigma}$. 

Following Dobrushin and Schlosman [1] we introduce the 
$\underline{\hbox{ground}}$ $\underline{\hbox{states}}$ 
simply as the Gibbs states at $\beta=\infty$; 
i.e. as the Gibbs states determined by the specification 
$$
\mu^{\infty H}_V(\sigma_V\mid{\bar\sigma})=
\lim_{\beta\to\infty}\mu^{\beta H}_V(\sigma_V\mid{\bar\sigma})
$$
Clearly
$$
\mu^{\infty H}_V(\sigma_V\mid{\bar\sigma})=
\cases{1/|M_V({\bar\sigma})| &if\quad 
$\sigma_V\in M_V({\bar\sigma})$ \cr
0 &if\quad $\sigma_V\not\in M_V({\bar\sigma})$ \cr}
$$
where 
$$
M_V({\bar\sigma})=\{{\tilde\sigma}_V\mid 
H_V({\tilde\sigma}_V\mid{\bar\sigma})=\inf_{\sigma_V}
H_V(\sigma_V\mid{\bar\sigma})\}
$$
is the set of 
$\underline{\hbox{ground configurations in }V
\hbox{ under the b.c. }\bar\sigma}$. 
A ground state will be called $\underline{\hbox{rigid}}$ if it
is supported by a single configuration $\sigma\in\Omega$, 
i.e. if it is a Dirac measure $\delta_\sigma$ on $\Omega$. 
However, the ground states may often be measures 
supported by a large set of configurations. 
Examples of such $\underline{\hbox{random ground states}}$ 
are met e.g. for the Ising antiferromagnet on triangular 
or FCC lattices.
We notice that even the problem of describing all the extremal 
periodic ground states is often non trivial. 
Let us mention in this connection the case of the three state 
Potts antiferromagnet on a square or cubic lattice which is 
still open.
An attempt to clarify it by mapping the ground states onto 
the Gibbs states of equivalent ferromagnetic models at a 
particular temperature was made in [2]. 

Of course, an important problem is to distinguish which 
ground states are $\underline{\hbox{stable}}$ in the sense 
that there exist Gibbs states at low temperature that are
``near'' to the ground states in question. 
Some particular cases of periodic rigid ground states 
are covered by the Pirogov-Sina{\"\i} theory [3] and a
a class of non translation invariant rigid ground states
was tackled in [4].
A general criterium for the stability of rigid ground states 
has been conjectured by Dobrushin and Shlosman [1].
However, no theory of stability exists for random 
ground states, although some statements about the thermodynamics 
involving such ground states were proven in a work by 
Aizenman and Lieb [5] about the third thermodynamical principle. 

When probing Gibbs states at low temperatures a useful 
notion may be thet of $\underline{\hbox{weak ground states}}$ 
[1] describing the effect of small perturbation added to the 
hamiltonian. 
Namely, considering an additional finite range interaction 
$\{{\tilde\varphi}_\Lambda\}$ and the corresponding 
hamiltonian ${\tilde H}$, a weak ground state 
(corresponding to the ``direction'' ${\tilde H}$) 
is a Gibbs state with the specification 
$$
\mu_V^{\infty H,{\tilde H}}(\sigma_V\mid{\bar\sigma})
=\lim_{\beta\to\infty}
\mu^{\beta(H+{\tilde H}/\beta)}_V(\sigma_V\mid{\bar\sigma})=
$$
$$\cases{
={{\displaystyle
\exp-{\tilde H}_V(\sigma_V\mid{\bar\sigma})}\over
{\displaystyle
\sum_{\sigma_V\in M_V({\bar\sigma})}
\exp-{\tilde H}_V(\sigma_V\mid{\bar\sigma})}}
&\quad if\quad $\sigma_V\in M_V({\bar\sigma})$ \cr\cr
=0 &\quad if\quad $\sigma_V\not\in M_V({\bar\sigma})$ \cr}
$$

Let us now inspect two particular exemples with an interesting 
structure of weak ground states. 
The case of periodic ground states will be discussed for an 
Ising antiferromagnet in an external magnetic field [1,6], 
while our main exemple, the ground states describing an 
interface and its roughening, 
will be discussed for an Ising model on a BCC lattice. 

Considering the Ising antiferromagnet on a square lattice 
with a nearest neighbour (n.n.) coupling $J$ and an external 
field $h$, one easily shows that, for $|h|<4|J|$ 
there are the two customary antiferromagnetic rigid 
ground states 
(stable according to, say the Pirogov-Sina{\"\i} theory). 
Inspecting now the border points, say $h=4|J|$, 
we get random ground states living on the set of all configurations 
for which no nearest neighbours are occupied by a pair 
of minus spins. 
Following [1,6] we may look at the lattice sites with 
minus spins as is they were occupied by a particle and thus 
we may equivalently think of a hard core lattice gas. 
Considering now the limit $\beta\to\infty$ along the lines 
$h=4|J|+\mu/\beta$ as shown in Fig.\ 1, i.e. 
a weak ground states in the direction $\tilde H$ 
given by the external field $\mu$, 
we get a hard core lattice gas with a chemical potential 
$-2\mu$, which is expected to have a critical value 
$\mu_c\sim -(1/2)\ln3.8$ (for a rigorous estimate see [6]). 
This suggest the phase diagram of Fig.\ 1.


\setlength{\unitlength}{1mm}
\begin{center}
\begin{picture}(110,40)

\put(50,0){\line(0,1){40}}
\put(0,0){\line(1,0){107}}
{\thicklines\bezier{300}(10,0)(20,25)(50,25)
\bezier{300}(50,25)(75,25)(90,0)}
\put(90,0){\line(-3,5){18}}
\put(74,31){$\mu_c$}

\put(90,0){\line(1,3){3}}
\put(98,24){\vector(-1,-3){5}}
\put(90,0){\line(1,1){7}}
\put(105,15){\vector(-1,-1){8}}
\put(107,15){$\mu$}
\put(90,0){\line(-2,1){10}}
\put(70,10){\vector(2,-1){10}}
\put(52,42){$T$}
\put(109,-2){$h$}
\put(5,-6){$-4|J|$}
\put(87,-6){$4|J|$}
\put(36,10){$\matrix{\hbox{Two ordered}\cr \,\hbox{phases}\hfill\cr}$}

\end{picture}
\end{center}

\bigskip

\centerline{Fig. 1\hphantom{000}}

\bigskip


Finally, let us consider the case of
an Ising model on a BCC lattice with a n.n. ferromagnetic coupling 
$J_0>0$ and a n.n.n. coupling $J$. 
Let us stress right now that we have in mind an isotropic model.
Whenever $J>-(2/3)J_0$ there are two stable, rigid, translation 
invariant ground states of constant magnetization. 
We shall enforce the (100) interface between these two 
phases by taking a b.c. ${\bar\sigma}$ with ${\bar\sigma}_i=+1$ 
if $i=(i_1,i_2,i_3)$ is a lattice site with $i_1\le0$ 
and ${\bar\sigma}_i=-1$ if $i_1<0$. 
For $J>0$ it is easy to show that such b.c. leads to 
a rigid ground state supported by ${\bar\sigma}$ itself.
Using the method of Dobrushin [8] as generalized in [4] 
or the method of van Beijeren [9], one may prove that this 
ground state is stable; actually one gets the existence 
of the corresponding Gibbs state with a rigid interface in
all the region shaded in Fig.\ 2 (with 
$\alpha_0=(1/2)\ln(1+\sqrt{2})$) denoting the critical value 
for the Ising model on a square lattice. 
A more interesting situation is obtained for $J=0$. 
Let us consider right away the weak ground states 
corresponding to the directions ${\bar H}$ yielded by a 
n.n.n. interaction of the form $J=\alpha/\beta$, 
i.e. the weak ground states
obtained along the lines shown in Fig.\ 2. 
One easily observes that the configurations in
$M_V({\bar\sigma})$ contain interfaces with no overhands. 
Actually these configurations are exactly those considered 
by van Beijeren in his body centered solid-on-solid 
(BCSOS) model [10]. 
Without going into the details [7] we may refer to his 
results to get a description of our weak ground states 
in terms of a six vertex model with the weights 
$\omega_1=\omega_2=\omega_3=\omega_4=e^{-\alpha}$, 
$\omega_5=\omega_6=1$. 
If $\alpha>\alpha_R=(1/2)\ln2$, 
the six-vertex model is in the ferroelectric phase 
and the interface is rigid; 
if $\alpha<\alpha_R$, the results about the six-vertex model 
are usually interpreted as describing a rough interface 
wich actually should mean that the corresponding infinite 
volume Gibbs state of the BCSOS model thus not exist 
and our weak ground state is translation invariant. 
Even though the above equivalence is exact only in the limit 
$\beta\to\infty$, one may expect that there is a curve 
$T_R(J)$ of roughening transitions as shown in Fig.\ 2. 


\setlength{\unitlength}{1mm}
\begin{center}
\begin{picture}(105,44)

\put(50,0){\line(0,1){40}}
\put(0,0){\line(1,0){100}}
\put(52,40){$kT/J_0$}
\put(102,-1){$J/J_0$}
{\thicklines\bezier{200}(50,0)(65,20)(80,30)}
\put(82,29){$T_R(J)$}
\put(50,0){\line(3,4){22}}
\put(72,31){$\alpha_R$}
\put(50,0){\line(-3,5){18}}
\put(28,30){$\alpha$}
\put(50,0){\line(2,1){37}}
\put(89,19){$\alpha_0$}
\put(50,0){\line(4,1){39}}
\put(50,0){\line(1,5){6.3}}
\multiput(52.5,0.5)(1,0){34}{.}
\multiput(54,1.5)(1,0){32}{.}
\multiput(56.5,2.5)(1,0){30}{.}
\multiput(58,3.5)(1,0){28}{.}
\multiput(60.5,4.5)(1,0){26}{.}
\multiput(62,5.5)(1,0){24}{.}
\multiput(64.5,6.5)(1,0){22}{.}
\multiput(66,7.5)(1,0){20}{.}
\multiput(68.5,8.5)(1,0){18}{.}
\multiput(70,9.5)(1,0){16}{.}
\multiput(72.5,10.5)(1,0){14}{.}
\multiput(74,11.5)(1,0){12}{.}
\multiput(76.5,12.5)(1,0){10}{.}
\multiput(78,13.5)(1,0){8}{.}
\multiput(80.5,14.5)(1,0){6}{.}
\multiput(82,15.5)(1,0){4}{.}
\multiput(84.5,16.5)(1,0){2}{.}

\end{picture}
\end{center}

\bigskip

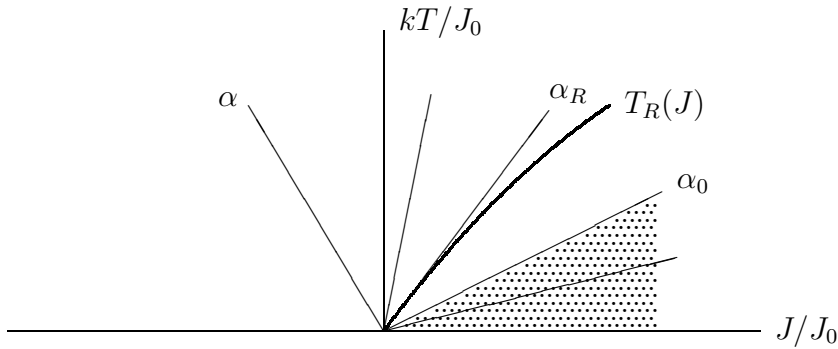
\centerline{Fig. 2}

\bigskip


Let us mention that one may investigate also the ground states 
with other b.c. ${\bar\sigma}({\vec k})$
corresponding to general inclined interfaces $(k_1,k_2,k_3)$ 
with normal $\vec k$. 
It turns out that for both positive and negative $J$ 
the ground state corresponding to an interface (110) is rigid 
and stable while the ground state corresponding to an interface 
(111) is translation invariant and the interface is rough. 
To see this fact one may (for any $J$ in an interval around 
zero) use a sililar equivalence as above to the triangular 
Ising solid on solid (TISOS) model and then refer to the 
analysis of this model by Nienhuis, Hilhorst and Bl{\"o}te 
[11] who solved it exactly. 
When one of the conditions 
$-k_1\le k_2+k_3\le k_1$, 
$-k_2\le k_3+k_1\le k_2$ or 
$-k_3\le k_1+k_2\le k_3$ is fulfilled the interface  
corresponding to the b.c. ${\bar\sigma}({\vec k})$ is again 
described in terms of a BCSOS model with the appropriate b.c. 
which is, in its turn, equivalent to a six vertex model 
with fixed polarizations. 
For normals $\vec k$ in the complementary region 
the corresponding interface may be  
described in terms of a TISOS model. 

To conclude let us comment about the connection of the 
roughening transition in the above sense and the facet formation 
in the equilibrium shape of a crystal (a droplet in the Ising model). 
One expects (see for instance [12]) that the roughening transition 
corresponds to the rounding of facets while a rigid interface 
associated to the b.c. ${\bar\sigma}({\vec k})$ would imply 
the presence of a cusp in the corresponding direction 
in the graph of the surface tension as a function of $\vec k$, 
and by the Wulff construction give rise to a plane facet. 
In fact following Bricmont, El Mellouki and Fr{\"o}hlich [12] 
one may use correlation inequalities to prove the existence 
of a cusp for the (100) facet in the shaded region of Fig.\ 2 
and for the (110) facet if $T$ is small enough and $J\ge 0$. 
In what concerns the rounding of the facets some insight 
may be obtained by introducing the free energy 
$$
f_V({\vec k})=S_V^{-1}\ln Z^{\rm SOS}(S_V,{\vec k}) 
$$ 
of the appropriate SOS model and b.c. associated as explained 
above to the b.c. ${\bar\sigma}({\vec k})$ of the Ising model 
in the volume $V$ ($S_V$ denoting the area of the interface 
$(k_1,k_2,k_3)$ let $e_V({\vec k})$ denote the energy of 
the corresponding ground state per unit area. 
Then, for finite volumes, one may show that the surface 
tension of the Ising model 
$$
\tau_V({\vec k})=S_V^{-1}\ln\big(Z(V,{\vec k})/Z^+(V)\big)
$$
behaves asymptoticaly with $T\to0$ as 
$$
\tau_V({\vec k})=e_V({\vec k})+kTf_V({\vec k})
$$
Supposing that this is correct also in the thermodynamic limit, 
the function $\tau_V({\vec k})$ con in principle be computed 
using the equivalence of the SOS models with b.c. with 
exact solvables models with fixed polarizations and from it 
get quantitative information on the equilibrium crystal 
shape as a function of the coupling constants. 

\underline{Note:}

The present text, 
publised in 
{\it VIIIth International Con\-gress on Mathematical Physics}, 
M. Mebkhout and R. S{\' e}n{\' e}or editors, 
World Scientific, Singapore, 1987 
(ISBN 9971-50-208-9), pp.\ 331--337, 
is our contribution to this conference,  
held in Marseille, France,  July 16--25, 1986.

This congress, 
organized by the 
International Association of Mathematical Physics, 
belongs to a series, that 
begun in 1972 at Moscow and has been pursued 
in 1974 at Warsaw, 
1975 Kyoto, 
1977 Roma, 
1979  Lausanne, 
1981 Berlin, and 
1983 Boulder. 
After Marseille, the congress took place
at Swansea, in 1988. 


\bigskip
\bigskip


\centerline{\bf REFERENCES}

\bigskip

\def\bib#1{\noindent[#1]\hphantom{0}}
\def\bibi#1{\noindent[#1]}

\parskip=4pt

\bib{1}
R.L. Dobrushin and S. Shlosman, 
Soviet Scientific Reviews {\bf C5}, 54 (1985) 

\bib{2} 
R. Koteck{\'y}, 
Phys. Rev. {\bf B31}, 3088 (1985)

\bib{3} 
S.A. Pirogov and Ya.G. Sina{\"\i}, 
Theor. Math. Phys. {\bf 25}, 1185 (1975);\break  
\hphantom{[00] }{\bf 26}, 39 (1976). 
For a review see Ya.G. Sina{\"\i}, ``Theory of Phase 
Transi-\break
\hphantom{[00] }tions; Rigorous Results'', 
(Pergamon, New York, 1982). 

\bib{4}  
P. Holick{\'y}, R. Koteck{\'y} and M. Zahradn{\'\i}k,
Rigid interfaces for lattice\break  
\hphantom{[00] }models at low temperatures, 
Preprint, Marseille, 1986.

\bib{5}  
M. Aizenman and E. Lieb, 
J. Stat. Phys. {\bf 24}, 279 (1981)  

\bib{6} 
R.L. Dobrushin, J. Kolafa and S. Shlosman, 
Commun. Math. Phys.\break  
\hphantom{[00] }{\bf 102}, 89 (1985)   

\bib{7} 
R. Koteck{\'y} and S. Miracle-Sole,
Phys. Rev. {\bf B34}, 2049 (1986)

\bib{8}
R.L. Dobrushin, 
Theory Prob. Appl. {\bf 17}, 582 (1972) 

\bib{9}
H. van Beijeren, 
Commun. Math. Phys. {\bf 40}, 1 (1975)   

\bibi{10}
H. van Beijeren, 
Phys. Rev. Lett. {\bf 38}, 993 (1977)

\bibi{11}  
E. Nienhuis, H.S. Hilhorst and H.W.J. Bl{\"o}te, 
J. Phys. {\bf A17}, 3559 (1984) 

\bibi{12}  
J. Bricmont, A. El Mellouki and J. Fr{\"o}hlich, 
J. Stat. Phys. {\bf 42}, 743\break 
\hphantom{[00] }(1986)  

\noindent 
{\it see also}, 

\bibi{13} 
R. Koteck\'y and S. Miracle-Sole, 
J. Stat. Phys. {\bf 47}, 773--799 (1987).

\end{document}